\documentclass[sigconf]{acmart}
\AtBeginDocument{%
  \providecommand\BibTeX{{%
    \normalfont B\kern-0.5em{\scshape i\kern-0.25em b}\kern-0.8em\TeX}}}

\copyrightyear{2021} 
\acmYear{2021} 
\setcopyright{rightsretained} 
\acmConference[CUI '21]{CUI 2021 - 3rd Conference on Conversational User Interfaces}{July 27--29, 2021}{Bilbao (online), AA, Spain}
\acmBooktitle{CUI 2021 - 3rd Conference on Conversational User Interfaces (CUI '21), July 27--29, 2021, Bilbao (online), AA, Spain}
\acmDOI{10.1145/3469595.3469620}
\acmISBN{978-1-4503-8998-3/21/07}

\acmConference[Conversational User Interfaces 2021]{CUI'21}{July 27--29}{Online}
\acmBooktitle{3rd Conference on Conversational User Interfaces (CUI'21),
  July 27--29, 2021}

\begin{document}

\title{StockBabble: A Conversational Financial Agent to support Stock Market Investors}

\author{Suraj Sharma}
\authornote{Both authors contributed equally to this research.}

\author{Joseph Brennan}
\authornotemark[1]

\author{Jason R.C. Nurse}

\email{J.R.C.Nurse@kent.ac.uk}
\orcid{0000-0003-4118-1680}

\affiliation{%
  \institution{University of Kent}
  \streetaddress{School of Computing}
  \city{Canterbury}
  \state{Kent}
  \country{UK}
  \postcode{CT2 7NF}
}

\renewcommand{\shortauthors}{Sharma, Brennan, Nurse}

\begin{abstract}
We introduce StockBabble, a conversational agent designed to support understanding and engagement with the stock market. StockBabble’s value and novelty is in its ability to empower retail investors -- many of which may be new to investing -- and supplement their informational needs using a user-friendly agent. Users have the ability to query information on companies to retrieve a general and financial overview of a stock, including accessing the latest news and trading recommendations. They can also request charts which contain live prices and technical investment indicators, and add shares to a personal portfolio to allow performance monitoring over time. To evaluate our agent's potential, we conducted a user study with 15 participants. In total, 73\% (11/15) of respondents said that they felt more confident in investing after using StockBabble, and all 15 would consider recommending it to others. These results are encouraging and suggest a wider appeal for such agents. Moreover, we believe this research can help to inform the design and development of future intelligent, financial personal assistants.

\end{abstract}




\maketitle

\section{Introduction}
\label{sec:introduction}
The current volatility of the stock market coupled with more access to time and disposable income (due to lack of traveling and shopping opportunities, for instance) has resulted in increased public interest in the financial stock market and in stock investment. Some studies report that as many as three in four younger individuals (millennials and generation Z) were planning to invest within the year \cite{rana_2021}.  Retail investors (such as individual or non-professional investors), in particular, are emerging as a key investor group. Moreover, as has been witnessed with the ongoing Reddit r/wallstreetbets versus Hedge Funds saga, these individuals have even acted to challenge longstanding institutional investors \cite{andersen_2021}. 

Non-professional individual investors as a user group come in many varieties, from novice traders to sophisticated day traders. One significant challenge that new and existing retail investors face is a lack of informational and interactive learning resources when compared with their institutional counterparts. For new investors, a lack of skill is often apparent \cite{martin_wigglesworth_2021,eaton2021zero} and can materialise in a limited understanding of financial jargon, technical stock performance charts, chart patterns across time, and trading decisions. These issues become more significant considering that such investors often face more direct financial risk because they are investing capital drawn from their own disposable income or savings. 

In this paper, we aim to support this emerging retail investor group through the proposal of a novel conversational agent capable of facilitating engagement with the financial stock market. The name of this agent is StockBabble. The motivation behind our agent is twofold. Firstly, we seek to empower retail investors and supplement their informational needs through the use of an interactive agent environment. Such agents have been applied in various related domains with good success \cite{jain2018evaluating}. StockBabble is designed to support a wide range of queries about the stock market, thus facilitating knowledge building, and it can also provide live news and performance statistics to inform decision-making on trades. 

\begin{figure*}[ht]
\centering
\includegraphics[scale=0.30]{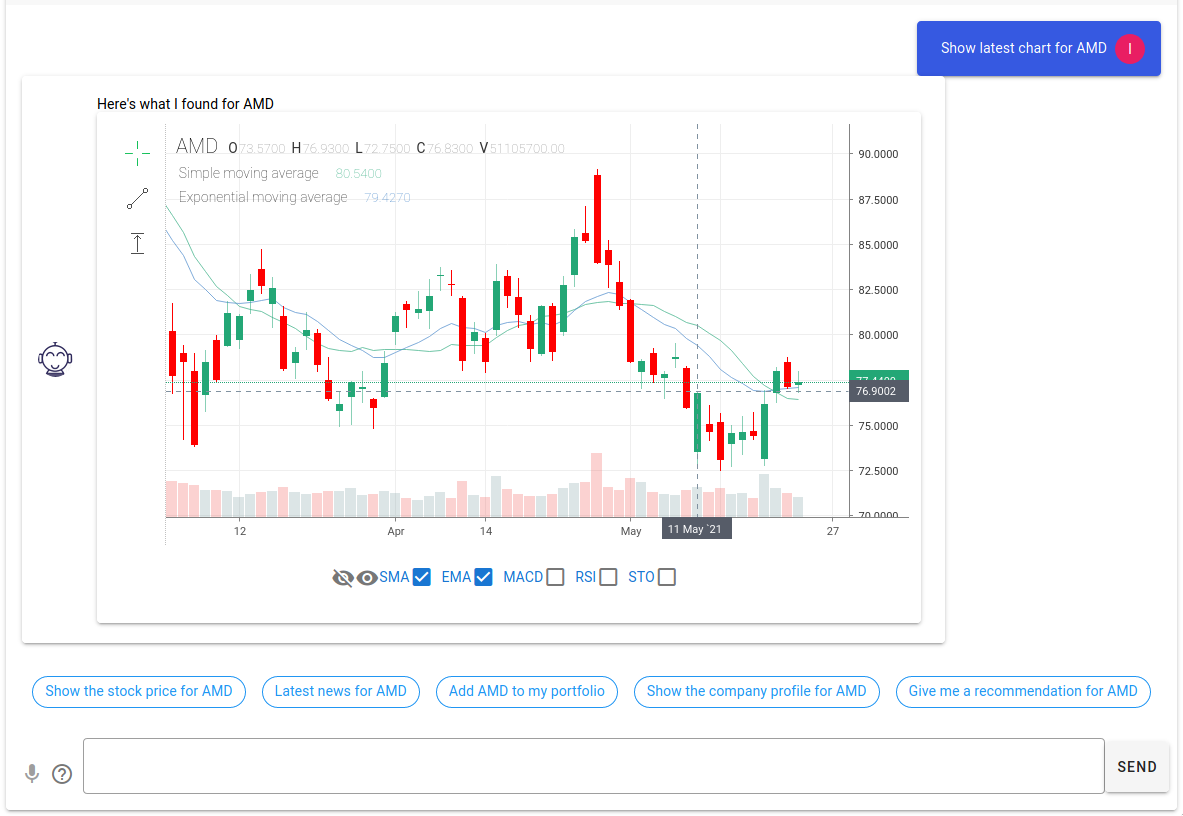}
\caption{An example interactive company chart within the StockBabble interface} 
\label{fig:1}
\Description[A screenshot of a chart showing the stock performance of AMD over time]{A screenshot of an interactive chart included within StockBabble, which shows the stock performance of any organisation (here, it is showing AMD). This chart presents a period of five months on the x axis, the price of AMD on the y axis, and various trend lines showing peaks and drops.}
\end{figure*}

The second motivation for our work is the lack of research into agents in the Conversational User Interface (CUI) community for supporting retail investors in the stock market. Wen \cite{wen2018conversational} and Lauren \& Watta \cite{lauren2019conversational} are two key articles in this domain, and while both are noteworthy, they primarily target prediction and forecasting – these are only a subset of the needs of new investors and make assumptions about users' understanding of the complexities of the stock market. They also lack specific features such as portfolio, terms and interactive charts, and are limited by their implementation platforms (i.e., LINE and the SlackAPI). We aimed to produce an agent where such limitations in scope can be avoided, and there is more freedom to design and prototype how each component can be engaged with by the end user.

\section{The StockBabble Agent}

\subsection{Functionality}
StockBabble is a conversational agent focused on empowering retail investors. It allows them to build their knowledge about the stock market and informs their decision-making through the use of a responsive chatbot environment. In many ways this agent disrupts the conventional means of accessing practical stock market information which is, at least for retail investors, traditionally based on directed browsing of a disparate set of websites and trading platforms/apps. Instead, we provide an agent capable of performing a series of common stock market look-ups and investment tasks. 

\begin{figure}[h]
  \includegraphics[scale=0.30]{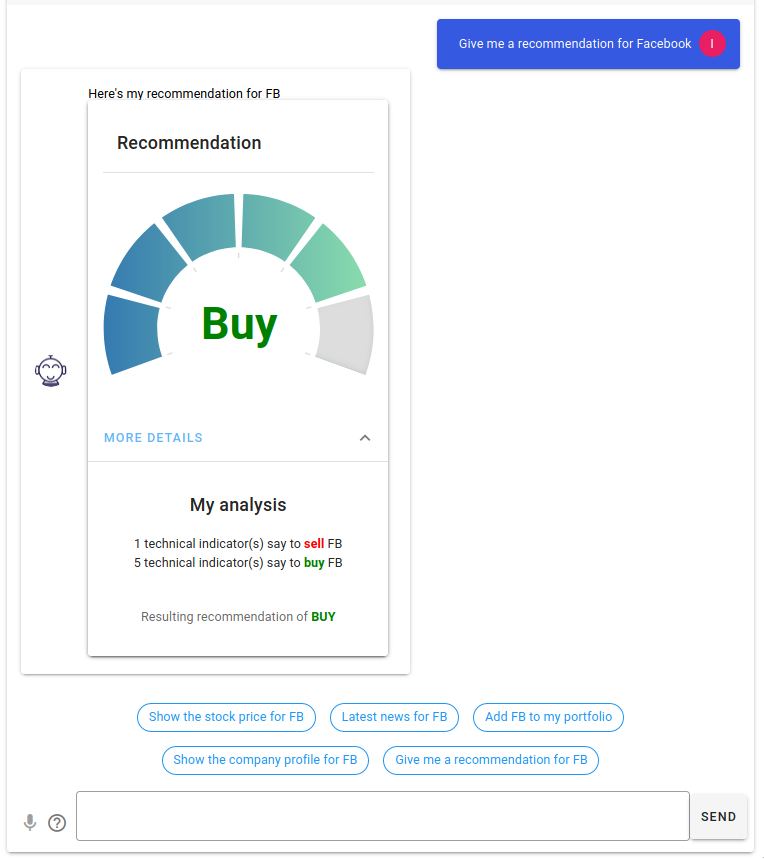}
\caption{An example recommendation that StockBabble produces based on six indicators, namely, SMA20, EMA20, EMA50, RSI, STO and MACD}
\Description[A screenshot showing a stock recommendation based on six indicators]{A screenshot showing a half circle with six areas aligned to the six stock performance indicators. Each area is coloured according to its prediction, i.e., buy or sell, and these are also presented in text form below the half circle. The aggregated indicators are then used to present a resulting recommendation, here, a Weak Buy for Facebook stock.}
\label{fig:2}
\end{figure}

The primary use cases that StockBabble targets are:

\begin{itemize}
  \item Knowledge building about the stock market and investments generally. This use case is grounded in the reality that many individuals may be new to investing and therefore there is value in an environment which is responsive and able to support their learning. StockBabble would therefore allow exploration of topics such as how the stock market works, as well as provide insights into key investment terms and jargon (e.g., what is a stock or how does an individual invest?). 
  
  \item Research and analysis of companies to help inform decision-making and investments. StockBabble is able to provide live news about organisations; a feature which we present in a timeline as it enables investors/users to view the latest information on stocks in order of time. There is also a stock price functionality that allows the user to ask for the latest stock price information for a company. This information is presented in an interactive chart, similar to what is seen on other trading platforms like TradingView\footnote{https://www.tradingview.com/chart/} or Trading212\footnote{https://www.trading212.com/en/Technical-Analysis}. Figure~\ref{fig:1} shows an example of the interactive chart with technical overlay data on the stock's performance. Additionally, we incorporated company profiles as a way in which users can review a complete overview of a company's financial and general business processes; this includes a range of information from where the business' head office is located, to the annual dividends the company distributes.  
  
  \item Stock portfolio and recommendation support is enabled through the features above, but StockBabble also contains a dedicated portfolio component to simulate how well stocks may perform before users execute on trades in the real world. Not only does the portfolio act as an ideal educational investment tool, but it also allows for end users to formulate their own investment strategies over time. To complement this, StockBabble is able to produce `buy' or `sell' recommendations based on a number of common technical stock indicators such as the Simple Moving Average (SMA) and Exponential Moving Average (EMA)~\cite{ig2019trade}. If the technical indicators pass a threshold, it represents either a buy or a sell. We tally the indicators to produce an overall recommendation which is then presented to the user. Figure \ref{fig:2} presents an example of this component/response, including the detailed indicators to justify the recommendation; this screenshot represents the final agent interface after updates based on user study feedback. Indicators are also integrated into overlays on the performance charts in StockBabble to assist the user in making a trading decision; these can be seen at the bottom of Figure~\ref{fig:1} as a toggle button for each indicator. The other indicators included are Relative Strength Index (RSI), Moving Average Convergence Divergence (MACD) and Stochastic Oscillator (STO)~\cite{ig2019trade}.

\end{itemize}

To improve engagement with StockBabble through the various use contexts above, we applied several design principles grounded in research~\cite{jain2018evaluating}. Firstly, we adopted a prompting technique where the agent would suggest the next question based on a user's previous question. For instance, if the user asks ``What is the stock price of Facebook'', the chatbot would first answer this question and then suggest additional follow-up queries, such as ``Show the company profile for Facebook'', or ``Give me a recommendation for Facebook''. Some examples of this can be seen in Figure~\ref{fig:1}. These could be accessed by the user to provide a more natural conversation flow and would seek to avoid known issues~\cite{luger2016like} related to poor agent responses and engagement. This button-type interface, alongside the textual input, is also recommended by research~\cite{jain2018evaluating} as it helps to make the conversation more natural and engaging. 

Additionally, users can interact with StockBabble by typing a message and pressing send, or by using voice input. Voice input is an increasingly popular way to engage with agents (due to agents such as Alexa, Siri and Google Assistant) and can provide users with another interaction mode~\cite{belennurse2020cui}; these factors motivated its inclusion.  Once voice recognition is enabled (by clicking the microphone -- see bottom left of Figure~\ref{fig:1}), a user can begin speaking and their words will be populated in the message box. We have adopted this approach for flexibility and to allow the text to be corrected if there are any errors before submission. The bottom of Figures~\ref{fig:1} and \ref{fig:2} depict this data entry area. Lastly, due to the size of some of the visual components which StockBabble produces (e.g., stock charts, recommendation information, and portfolio data), we automatically minimise the prior components once another question has been asked. For example, if the user asks for a chart for a company followed by the news, the chart component would be minimized. Users can easily maximise it again by clicking on it. This approach to displaying information helps to support a better conversation flow, promote a cleaner stream of utterances, and reduce the bulkiness of the dialogue. 

\subsection{Design, Development and Technical details }

\begin{figure}[ht]
\centering
\includegraphics[scale=0.80]{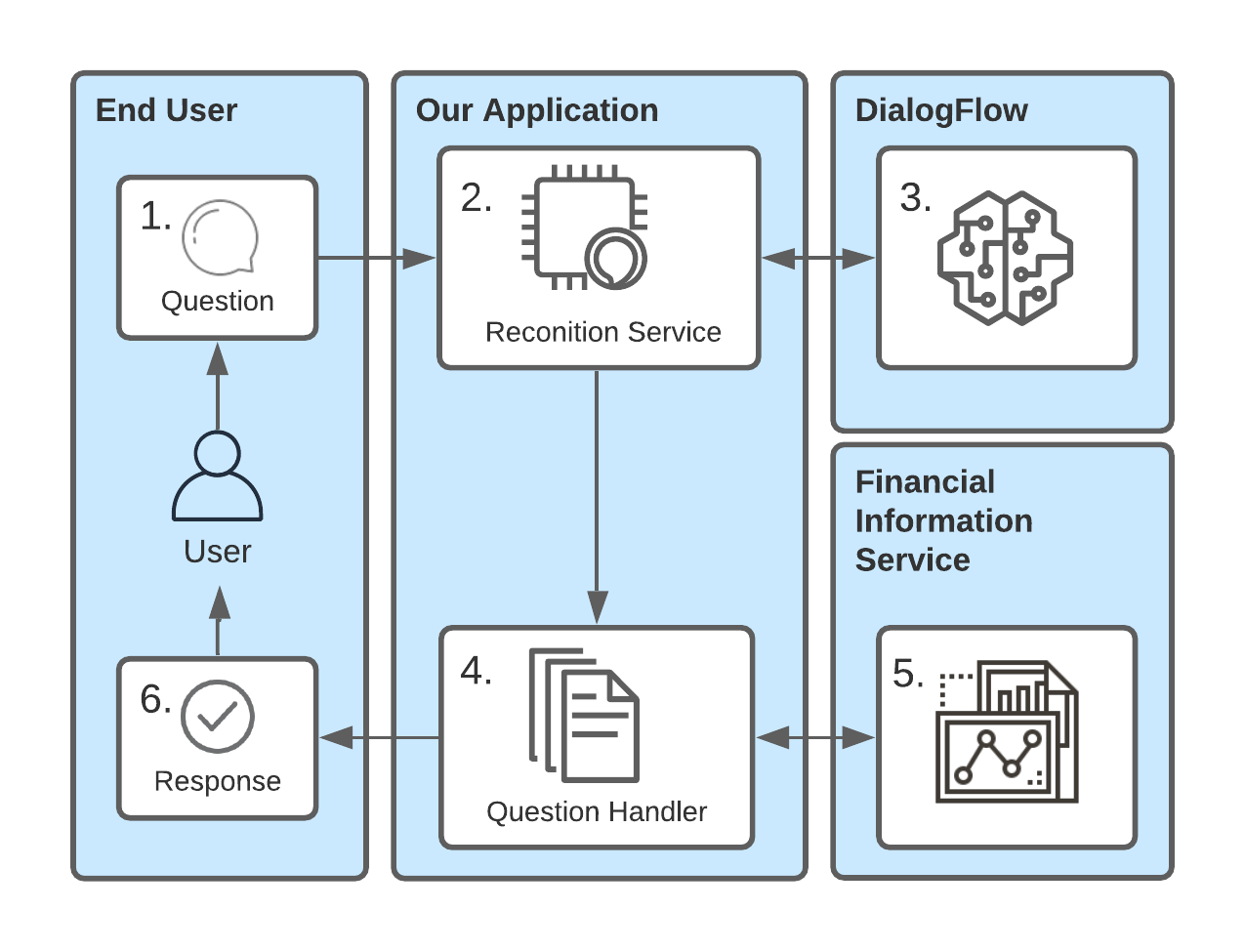}
\caption{An overview of StockBabble's architecture and how a user query is processed} 
\label{fig:3}
\Description[The high level architecture of StockBabble and a presentation of the main activities]{The high level architecture of StockBabble, presenting the six main activities: (1) The question is posed by the user, (2) a recognition service accepts the question and passes it to DialogFlow, (3) DialogFlow interprets the question and its key components, (4) these components are passed to the question handler by DialogFlow (via the recognition service), (5) any necessary financial information is queried using another API, and (6) the response is presented to the user.}
\end{figure}

The design and development of StockBabble was carried out part-time over a period of five months. To create the agent, we drew requirements from the use cases of a retail investor (both key needs and challenges) and discussions with investors, which can generally be seen in the functionalities defined above. Our design began with low fidelity prototypes which were iterated over through brainstorming sessions with persons with varying stock market and investing experience. Once we had a stable set of prototype designs, we began development of the StockBabble conversational agent. We built StockBabble as a web app available on desktop and mobile platforms. After registration (registration was needed to allow user portfolios to persist), the user would be presented with a conversational interface which allows them to interact with the agent using voice and text. The front-end application, which is displayed to the user, communicates with a back-end API which is used for the processing of questions and retrieval of information. To provide more detail, Figure \ref{fig:3} shows the users journey upon asking a question.

StockBabble's front-end application was written using the VueJs\footnote{https://vuejs.org/} JavaScript framework and Web Speech API's speech recognition tool to implement voice input\footnote{https://developer.mozilla.org/en-US/docs/Web/API/Web\_Speech\_API}. The backend API was written using NodeJs\footnote{ https://nodejs.org/} with the Express framework for routing. When this API receives a user’s questions (e.g., “What’s the price of Amazon today?”), we send the request to our Google DialogFlow\footnote{  https://cloud.google.com/dialogflow}
agent, which assesses the content and returns what it recognised for further processing. DialogFlow acted as a pivotal component in meeting our requirements, particularly through its use of machine learning to recognise variations of expressions using training data. This meant that we did not have to initially manually add sentences to anticipate phrasing of queries and could concentrate on aggregating a suite of robust tools to provide the best agent platform/interface for retail investors.

We categorised the use cases generally into intents which are triggered from a user's expression. Intents allow us to understand what the user is asking. For example, if a user asked, ``Who is the CEO of Facebook'', our agent would recognise this question as the company's profile intent and within that, a question about who the company's CEO is. Or, if a query was posed about ``what stocks are'', the agent would categorise this question as a trading term intent. Our back-end (question handlers) would then deal with finding terms associated to the one requested to show related terms as well.

The assorted expressions within DialogFlow are further supported through the use of `entities' to ensure the intents never experience any crossover or disruptive overlap with the user’s input. Entities have allowed us to provide keyword reference values against user entries, this helps us to appropriately differentiate specific intents from each other. Once recognised, we can process the user's request using our API. The API includes a range of intent handlers. These handlers will fetch the required information, e.g., stock information, companies in portfolio, term definitions as well as recommendations. If Dialogflow categorises a question under the `StockInformation' intent, for example, our back-end API will use the `StockInformation' handler to fulfil the users question. Within this handler, a financial API is queried using the company name Dialogflow has recognised within the users question. Once financial information is retrieved we apply our technical indicators. A response is then constructed and presented on the front-end. 

\section{User Study}
To evaluate the usefulness, utility, and potential of StockBabble as a conversational financial agent, we conducted a user study with 15 participants with varying financial and stock market knowledge. This was a convenience sample based on individuals with the range of experience desired. This study would also allow us to gather feedback that could assist in the improvement of the agent platform and environment. The study involved a survey where participant were first asked questions related to themselves, their financial knowledge, if they had ever invested in a stock, and what they would expect from a financial chatbot application. After answering the initial questions, the user was provided with login details and instructions on how to use the agent. The users were encouraged to spend around 15 minutes using the agent. 
Once the participant had finished using StockBabble, we asked questions related to their experience, their overall impression, whether they learned anything from using the agent, what their primary mode of navigation was, what they found most useful, whether it met their expectations, and whether they would recommend it to others.

Overall, participants were very positive about StockBabble. 73\% (11) of our 15 participants said they felt more confident in investing after using StockBabble, 73\% (11) said it met their expectations, and all would consider recommending it to others. A few notable comments were: ``Very comprehensive recommendations'', ``It also provided a wide range of news articles about different topics to help understand the company's impact/performance better'', ``I didn't really know what to ask StockBabble so I found the suggestions helpful, for telling me what I needed to consider when thinking about Stocks'', and ``Lovely interface it made my experience even better!''. Beyond these points, one finding that was particularly interesting was that all participants used typing as their predominant mode of interaction as opposed to voice input. We implemented this feature assuming that speaking would be more natural, even though text input is more standard online. This is certainly an area to engage further in future work, and through qualitative study.

The study highlighted a few key areas for improvement which are important to increasing the utility of StockBabble and related agents. One example pertained to the technical ability of StockBabble and making the agent more intelligent; this was because in some cases it would not recognise certain questions and thus led to user frustration. This hints to the larger issues of accurate natural language processing and correct responses which are fundamental to the CUI community~\cite{yuksel2017brains}. In our case, we sought to address this concern by adding several new stock market-related training phrases within DialogFlow to capture a wider range of cases. 

Another feedback point pertained to modifying the recommendation component to explain how the recommendation was obtained, or as noted by a participant, ``Possible explanation as to why StockBabble has given that recommendation would be even better, especially for newer/more conservative traders.'' This can be interpreted more generally to highlight the importance of agents to be transparent and provide reasons behind certain guidance or suggestions; these actions can help to build user trust. This builds on similar discussions in the literature around the transparency of agents~\cite{hepenstal2019algorithmic}. To address this, we first reduced the amount of stock indicators shown by default when presenting a company chart. This was done to prevent overwhelming individuals with technical information. Once we did this, we enabled a drop down extension to the recommendation interface and showed how StockBabble produces its analysis by totaling the outputs of the indicators. The updated interface can be seen in Figure~\ref{fig:2}. 

\section{Demo}

The goal of our demonstration will be to introduce StockBabble to the  CUI community, both as a new agent but also as an idea for a new application context for such interfaces. As mentioned in Section~\ref{sec:introduction}, there is very little academic research in this domain and only recently are we witnessing finance-related agents take significant steps in industry~\cite{finchatbot}. The demo will begin with a short motivation for StockBabble and how it can be used by retail investors to significantly improve current practice. We then present the three main use cases assuming the role of Isabelle, a new retail investor, looking for technology companies to invest in. This will start with Isabelle asking StockBabble questions about the stock market and common trading jargon that she has seen being used on Reddit's r/wallstreetbets. She then begins to explore technology companies' (e.g., Apple, Amazon, Alphabet, AMD) profiles and recent stock-related news, before asking the agent for recommendations on whether she should consider investing. To make this session engaging and interactive, online attendees will be asked to help play the role of Isabelle. We will encourage individuals to suggest their own queries, and help to build a strong portfolio of technology stocks. We will also have an online version of StockBabble that attendees can access during the demo and conference. If any issues arise (e.g., StockBabble does not understand a query or returns an inappropriate response) we will use this as an opportunity to discuss these concerns with attendees in the context of chatbots generally and the implications for future stock-market CUIs.  

\section{Conclusions and Future Work}
As the stock market becomes more appealing to new retail investors, there is a growing opportunity to explore the use of conversational user interfaces that can break barriers of engagement and support better interactions. We set out to investigate such an agent which combines a natural language interface with various back-end technologies to provide stock market information to investors. From our research thus far, StockBabble has achieved this aim and provides a platform that allows users to converse with an agent for multiple investment-related purposes. StockBabble is able to support knowledge building about the stock market and investments generally, it can facilitate research and analysis of companies to inform decision-making on trades, and users can create a portfolio of stocks and receive buy/sell recommendations. 

StockBabble as a niche and innovative concept has the potential to further inform future financial intelligent personal assistants in the CUI community. Through more intricate stock analysis, user feedback, and future improvements, this concept could easily become commercialised. However, this can only happen if necessary improvements are made to its core infrastructure and functionality in collaboration with its primary end-user groups. We also note that there are regulatory considerations around who is legally allowed to provide financial advice to individuals and this will also need to be addressed. In the future, we will aim to further add to the functionality of StockBabble through more comprehensive use cases and extensive user studies. 

\section{Acknowledgements}
This work is funded by the UK Engineering and Physical Sciences Research Council (EPSRC) under the `A Platform for Responsive Conversational Agents to Enhance Engagement and Disclosure (PRoCEED)' research project (EP/S027297/1).

We would also like to thank everyone who participated in early use case discussions, in prototyping, and in evaluating StockBabble to help us build and improve the agent.

\bibliographystyle{ACM-Reference-Format}
\bibliography{main}

\end{document}